\documentclass{acm_proc_article-sp}

\toappear{}
\usepackage{balance}  % to better equalize the last page
\usepackage{graphics} % for EPS, load gr1aphicx instead
\usepackage{url}      % llt: nicely formatted URLs
\usepackage{graphicx}
\usepackage{caption}
\DeclareCaptionType{copyrightbox}
\usepackage{subcaption}
\def\etal{et~al.\,}
\usepackage{comment}
\usepackage[lined,boxed,commentsnumbered]{algorithm2e}
\usepackage{multicol}

\def\eg{\emph{e.g.}\,}

\usepackage[pdftex]{hyperref}
\usepackage[pdftex]{hyperref}
\hypersetup{
pdftitle={The Extreme Right Filter Bubble},
pdfauthor={Derek O'Callaghan, Derek Greene, Maura Conway, Joe Carthy, Padraig Cunningham},
pdfsubject={extreme right; YouTube; filter bubble; matrix factorization},
pdfkeywords={extreme right; YouTube; filter bubble; matrix factorization},
bookmarksnumbered,
pdfstartview={FitH},
colorlinks,
citecolor=black,
filecolor=black,
linkcolor=black,
urlcolor=black,
breaklinks=true,
}
\begin{document}

\title{The Extreme Right Filter Bubble}

% Note that submissions are blind, so author information should be omitted
\numberofauthors{5}
%\begin{comment}
\author{
  \alignauthor Derek O'Callaghan\\
    \affaddr{University College Dublin}\\
    \affaddr{Dublin 4, Ireland}\\
    \email{derek.ocallaghan@ucd.ie}\\
  \alignauthor Derek Greene\\
     \affaddr{University College Dublin}\\
    \affaddr{Dublin 4, Ireland}\\
    \email{derek.greene@ucd.ie}\\
   \alignauthor Maura Conway\\
   \affaddr{Dublin City University}\\
    \affaddr{Dublin 9, Ireland}\\
    \email{maura.conway@dcu.ie}
\and
   \alignauthor Joe Carthy\\
   \affaddr{University College Dublin}\\
    \affaddr{Dublin 4, Ireland}\\
    \email{joe.carthy@ucd.ie}\\
   \alignauthor P\'{a}draig Cunningham\\
   \affaddr{University College Dublin}\\
    \affaddr{Dublin 4, Ireland}\\
    \email{padraig.cunningham@ucd.ie}\\
}
%\end{comment}

\begin{comment}
\author{
  \alignauthor Author\\
    \affaddr{Institute}\\
    \affaddr{Address}\\
    \email{email}\\
  \alignauthor Author\\
     \affaddr{Institute}\\
    \affaddr{Address}\\
    \email{email}\\
   \alignauthor Author\\
   \affaddr{Institute}\\
    \affaddr{Address}\\
    \email{email}
\and
   \alignauthor Author\\
   \affaddr{Institute}\\
    \affaddr{Address}\\
    \email{email}\\
   \alignauthor Author\\
   \affaddr{Institute}\\
    \affaddr{Address}\\
    \email{email}\\
}
\end{comment}

% Teaser figure can go here
%\teaser{
%  \centering
%  \includegraphics{Figure1}
%  \caption{Teaser Image}
%  \label{fig:teaser}
%}

%\begin{comment}

\maketitle

%\DeclareCaptionType{copyrightbox}
\begin{abstract}

Due to its status as the most popular video sharing platform, YouTube plays an important role in the online strategy of extreme right
groups, where it is often used to host associated content such as music and other propaganda. In this paper, we develop a 
categorization suitable for the analysis of extreme right channels found on YouTube. By combining this  
with an NMF-based topic modelling method, we categorize channels originating from links propagated by extreme right Twitter accounts.
This method is also used to categorize related channels,  which are determined using results returned by YouTube's
related video service.  
%We identify the existence of a \emph{filter bubble}, whereby users who access an extreme right YouTube video are highly likely to be
We identify the existence of a ``filter bubble'', whereby users who access an extreme right YouTube video are highly likely to be
recommended further extreme right content.

\end{abstract}


%\end{comment}

%\terms{
%	Human Factors; Design; Measurement. 
%	If you choose more than one ACM General Term, 
%	separate the terms with a semi-colon.
%\\
%\textcolor{red}{If you choose more than one ACM General Term, 
%separate the terms with a semi-colon. See list of ACM terms at:
%\url{http://www.sheridanprinting.com/sigchi/generalterms.htm}.
%Optional section to be included in your final version.}
%}

%!TEX root = main.tex
\section{Introduction}

Social media platforms have become a key component in the online strategy of extremist political groups, as they provide direct access to
audiences, particularly youth, for radicalization and recruitment purposes \cite{ZwischenPropaganda2011,Conway:2008:JVA:1485445.1485462}.
YouTube's status as the most popular video sharing platform means that it is especially
useful to these groups. Extreme right content such as music and other associated propaganda, some of which is deemed illegal in certain
countries, is made freely available there, often for long periods of time \cite{SpiegelNaziRock2013}. In addition to hosting user-generated video content,
YouTube also provides  recommendation services, where sets of related and recommended  videos are  presented  to users, based on factors such
as co-visitation count and prior viewing history \cite{Davidson:2010:YVR:1864708.1864770}. 

In 2011, Pariser proposed the \textit{filter bubble} concept, which stated that recommendation services may limit the types of content
to which a user is exposed \cite{ParisierFilterBubble2011}. Recommendation in social media can therefore have the undesirable consequence 
of a user being excluded from information that is not aligned with their existing perspective, potentially leading to immersion within an
ideological bubble.
In this paper we show that this can happen when a user accesses an extreme right (ER) video or channel on YouTube. In studies on English
and German ER content, we show that users are likely to be recommended further ER content within the same category, or related ER content
from a different category. They are unlikely to be presented with non-ER content (Figure \ref{fig:summary}).

Our previous work found that Twitter is used by extreme right groups to propagate links to content hosted on thirdparty
sites, including YouTube  \cite{O'Callaghan:2013:UWS:2464464.2464495}. These links enabled us to retrieve video and uploader channel
information from YouTube, where subsequent analysis is focused upon \emph{channels} (synonymous with uploaders or accounts \cite{Simonet2013}), rather than individual videos. 
We use a topic modelling strategy to characterize these channels (see Section \ref{methodology}), where content linked to these channels
is then characterized in the same fashion. Our proposed scheme includes categories identified in the academic literature on the extreme
right, which are suited to the videos and channels found on YouTube. We find evidence that an extreme right filter bubble does indeed exist,
as shown by the results discussed in Section \ref{analysis}.

In Section \ref{relatedwork}, we provide a description of prior work on YouTube categorization and recommendation, along with research that
involved the categorization of online extreme right content. The retrieval of YouTube data based on links originating from extreme right
Twitter accounts is then discussed in Section \ref{data}. Next, in Section \ref{methodology}, we describe the methodology used for related channel ranking, topic identification, and channel categorization. Our investigation into the extreme right filter bubble can be found in Section
\ref{analysis}, where we focus on two data sets consisting primarily of English and German language channels respectively. An overview of
the entire process can be found in Figure \ref{fig:process}.

\section{Related Work}
\label{relatedwork}

\subsection{Video Recommendation}
\label{videorecommendation}

Video recommendation on YouTube has been the focus of a number of studies. For example, Baluja \etal suggested that
standard approaches used in text domains were not easily applicable due to the difficulty of reliable video labelling
\cite{Baluja:2008:VSD:1367497.1367618}. They proposed a graph-based approach that utilised the viewing patterns of YouTube users, which did
not rely on the analysis of the underlying videos. The recommendation system
 in use at YouTube at the time was discussed by Davidson \etal, where
sets of personalized videos were generated with a combination of prior user activity (videos watched, favorited, liked) and the traversal of
a co-visitation graph \cite{Davidson:2010:YVR:1864708.1864770}. In this process, recommendation diversity was obtained by means of a limited
transitive closure over the generated related video graph. Zhou \etal performed a measurement study on YouTube videos to determine
the sources responsible for video views, and found that related video recommendation was
the main source outside of the search function for the majority of videos \cite{Zhou:2010:IYR:1879141.1879193}. They also found that the
click through rate to related videos was high, where the position of a video in a related video list played a critical role. A similar
finding was made by Figueiredo \etal, where they demonstrated the importance of key mechanisms such as related videos in the attraction of
users to videos \cite{Figueiredo:2011:TOT:1935826.1935925}.

Turning to the task of YouTube video categorization, Filippova \etal presented a text-based method that relied upon metadata such
as video title, description, tags and comments, in conjunction with a pre-defined set of 75
categories \cite{Filippova:2011:IVC:2009916.2010028}. Using a bag-of-words model, they found that all of the text sources contributed to
successful category prediction. More recently, a framework for the categorization of video channels was proposed by Simonet, involving the use of semantic entities
identified within the corresponding video and channel profile text metadata \cite{Simonet2013}. Following the judgement that existing
taxonomies were not well-suited to this particular problem, a new category taxonomy was developed for YouTube content. Roy \etal
investigated both video recommendation and categorization in tandem, where videos were categorized according to topics built from Twitter
activity, leading to the enrichment of related video recommendation \cite{Roy:2012:SCT:2393347.2393437}. Video text metadata was used for
this process, and the topics were based on the categories proposed by Filippova \etal, in addition to the standard YouTube categories at
the time. Separately, they also analyzed diversity among related videos, where they found that there was a 25\% probability 
on average of a related video being from a different category.

\subsection{Extreme Right Categorization}
\label{ercategorization}

In the various studies that have analyzed online extreme right activity, certain differences can be observed among researchers
in relation to their categorization of this activity and associated organizations \cite{Blee2010}. Burris \etal proposed a
set of eight primarily US-centric categories in their analysis of a white supremacist website network; Holocaust Revisionists, Christian
Identity Theology, Neo-Nazis, White Supremacists, Foreign (non-US) Nationalists, Racist Skinheads, Music, Books/Merchandise
\cite{Burris2000}. A similar schema was used by Gerstenfeld \etal, which also included Ku Klux Klan and Militia categories
\cite{GerstenfeldHateOnline2003}. They also discussed the difficulty involved in the categorization of certain sub-groups, where a
general category (Other) was applied in such cases. These categories were adapted in separate studies of Italian and German extreme right
groups, where new additions included Political Parties and Conspiracy Theorists, while others such as Music and Skinheads were
merged into a Young category (Tateo \cite{JCC4:JCC410}, Caiani \etal \cite{doi:10.1080/13691180802158482}). Rather than focusing on
ideological factors, Goodwin proposed four organizational types found within the European extreme right milieu; political parties,
grassroots social movements, independent smaller groups, and individual `lone wolves'. Other notable categories include the Autonomous Nationalists identified within Germany
in recent years. These groups focus specifically on attracting a younger audience, where social media is often a critical component in this
process \cite{ZwischenPropaganda2011}.

The popularity of YouTube has led to its usage by extreme right groups for the
purpose of content dissemination. Its related video recommendation service provides a motivation for the current work to analyze the extent
to which a viewer may be exposed to such content. Separately, disagreements over the categorization of online extreme right activity suggests that a
specific set of categories may be required for the analysis of this domain.

%!TEX root = main.tex
\section{Data}
\label{data}

In our previous work, we investigated the potential
for Twitter to act as one possible gateway to communities within the wider online network of the extreme right \cite{O'Callaghan:2013:UWS:2464464.2464495}. Two data
sets associated with extreme right English language and German language Twitter accounts were generated, by retrieving profile data over an extended period
of time. We gathered all tweeted links to external websites, and used these to construct an extended network representation. In the
current work, we are solely interested in tweeted YouTube URIs. Data for the Twitter accounts were retrieved between June 2012 and May 2013, as limited by the Twitter API restrictions effective at the time. YouTube URIs found in tweets were analyzed to determine a set of channel (account)
identifiers that were directly (channel profile page URI) or indirectly (URI of video uploaded by channel) tweeted. All identified channels were included,
regardless of the number of tweets in which they featured. Throughout this work, we refer to these as \textit{seed channels}; 26,460
and 3,046 were identified for the English and German data sets respectively.

\begin{figure*}[!t]
    \centering
    \includegraphics[scale=0.69]{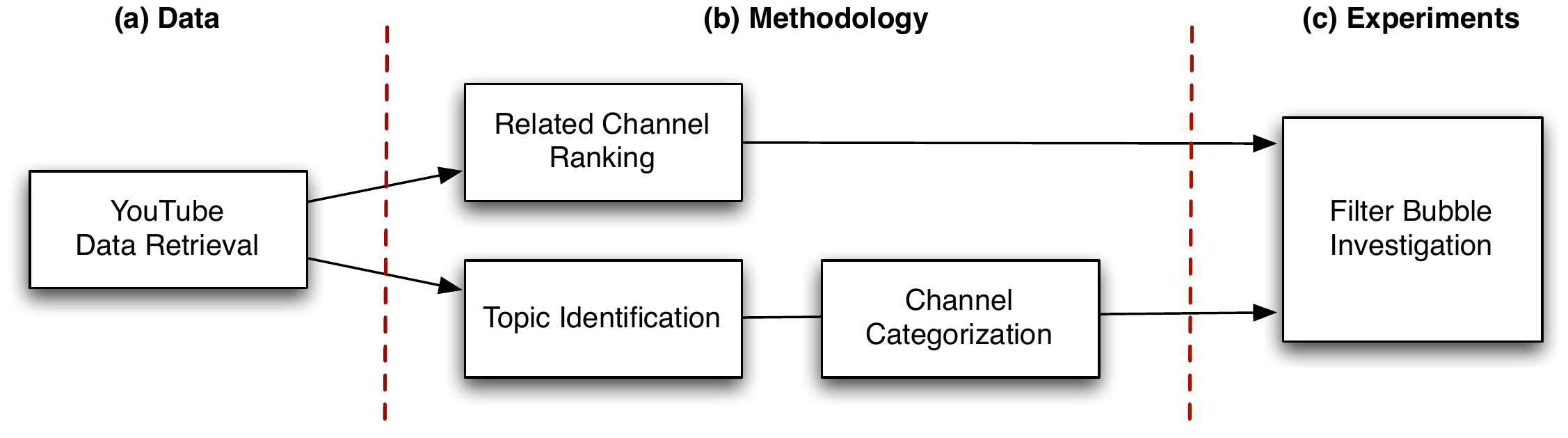}
	\caption{Overview of the process used to investigate the extreme right filter bubble on YouTube.}
    \label{fig:process}
\end{figure*}

In order to explore the filter bubble hypothesis we must first categorize YouTube channels. To do this, we use text metadata associated with
the videos uploaded by a particular channel, namely their titles, descriptions and associated keywords. Although user comments have been
employed in other work \cite{Filippova:2011:IVC:2009916.2010028}, they were excluded here. This decision followed an initial manual analysis
of a sample of tweeted videos, which found that comments were often not present, or had been explicitly disabled by the uploader. We
also excluded the YouTube ``category'' field as it was considered too broad to be useful in the extreme right domain. Using the YouTube API,
we initially retrieved the available text metadata for up to 1,000 of the videos uploaded by each seed channel, where the API returns videos
in reverse chronological order according to their upload time. In cases where seed channels and their videos were no longer available (\eg
the channel had been suspended or deleted since appearing in a tweet), these channels were simply ignored. 

In order to address the variance in the number of uploaded videos per channel, and to reduce the volume of subsequent data
retrieval, we randomly sampled up to 50 videos for each seed. For each video in this
sample, metadata values were retrieved for the top ten \emph{related videos} returned by the API. 
Using the default parameter settings, these videos appear to be returned in order of ``relevance'', as defined internally by YouTube,
 similar to the default behaviour of video search results feeds described in the API documentation (June 2013).
We refer to the corresponding uploaders as \textit{related channels}; 1,451,189
and 195,146 were identified for the English and German data sets respectively.
As before, we then retrieved the available text metadata for up to 1,000 videos uploaded by each unseen related channel, from which a random
sample of up to 50 videos was generated. 

For both seed and related channels, the corresponding uploaded video sample is used for
categorization; this is described in Section \ref{topicidentification}. 
Separately, we also note that the videos in question may have been uploaded at any time prior to retrieval, where these times are not
necessarily restricted to the period of either Twitter or YouTube data retrieval.

\section{Methodology}
\label{methodology}

Having retrieved the channel and video data from YouTube, the next steps involved ranking the related channels for each seed channel,
identifying latent topics associated with the uploaded videos, and using these to categorize both seed and related channels. This section
corresponds to part (b) of the process overview diagram found in Figure \ref{fig:process}.
 
\subsection{Related Channel Ranking}
\label{ranking}

For our analysis, it was first necessary to generate a set of related channel rankings for each seed channel,
according to the related rankings returned by the YouTube API for the sample of videos uploaded by the seed. We applied the SVD rank
aggregation method proposed by Greene~\etal to combine the rankings for each uploaded video into a single ranked set across all videos
for the seed in question, from which we select the top ranked related channels 
\cite{Greene:2013:PUG:2464464.2464471}. We restrict our focus to the top 10, given the impact of related video position on click through rate
\cite{Zhou:2010:IYR:1879141.1879193}. The aggregation process for each seed channel $s_i$ and its $n_i$ uploaded videos is as follows:

\begin{enumerate}
    \item For each uploaded video $v_{ij}$ ($j \in [1,n_i]$), generate a rank vector $rv_{ij}$ between the seed $s_i$ and its
    related channels using the retrieved YouTube API related video ranking. For channels related to seed videos other than $v_{ij}$, a rank
    of $(m_i + 1)$ is applied, where $m_i$ is the total number of channels related to $s_i$.
    \item Stack all $rv_i$ rank vectors as columns, to form the $m_i \times n_i$ rank matrix $R_i$, and normalise the columns of this matrix
    to unit length.
    \item Compute the SVD of $Ri$, and extract the first left singular vector. Arrange the entries in this vector in
    descending order to produce a ranking of all channels related to this seed. %Then select the top 10 ranked channels.
\end{enumerate}
From the ranking generated above, we then select the top 10 ranked related channels for each seed. These  channels are used for our filter bubble investigation in Section \ref{analysis}.

\subsection{Topic Identification}
\label{topicidentification}

For the purpose of channel categorization, we initially identified latent topics associated with the channels in both data sets, based on
their uploaded videos. Following the approach of Hannon~\etal \cite{hannon10twitter}, we generated a ``profile document'' for each seed and related
channel, consisting of an aggregation of the text metadata from their corresponding uploaded video sample, from which
a tokenized representation was produced.
All available terms in the channel documents were used, where the tokenization process involved the exclusion of URIs and normalization of
diacritics. We excluded a custom set of stopwords, such as official YouTube category terms, additional
YouTube-specific terms such as ``video'' and ``view'', and terms from multiple language stopword lists. Due to the frequent presence of
names in the data, we also excluded all identified first names.
As suggested by Fillipova \etal \cite{Filippova:2011:IVC:2009916.2010028}, terms were not stemmed due to the expectation of both noisy terms
and those of multiple languages. Low-frequency terms appearing in $<20$ channel documents were excluded at this point. 

\begin{table*}[!t]
\caption{Categories of extreme right YouTube content, based on common categorizations found in academic literature on extreme right
ideology.}
\begin{center}
\vskip -0.8em
\begin{tabular}{| l | p{10.7cm} | p{1.7cm} |}
\hline 
\emph{Category} & \emph{Description} & \emph{Source}
\\
\hline \hline  
Anti-Islam \hspace{1em} & Can include political parties (\eg Dutch \textit{PVV}) or groups such as the English Defence League
(EDL), which often describe themselves as ``counter-Jihad''.
& \cite{ZwischenPropaganda2011,GoodwinEDLCounterJihad2013}
\\
\hline 
Anti-Semitic & All types of anti-Semitism, regardless of association (existing literature tends to discuss this in relation to other
categories).
& \cite{Burris2000,JCC4:JCC410} \\ \hline 
Conspiracy Theory & Themes include New World Order (NWO), Illuminati etc. Not exclusively ER, but often related to Patriot in this context.
& \cite{ZwischenPropaganda2011,SPLCIdeology2013,JCC4:JCC410} \\
\hline 
Music & Includes any ER music such as Oi!, Rock Against Communism (RAC) etc. & \cite{ZwischenPropaganda2011,Burris2000,JCC4:JCC410}
\\ 
\hline 
Neo-Nazi & Nazi references, such as to Hitler, WWII, SS etc. & \cite{ZwischenPropaganda2011,Burris2000,GerstenfeldHateOnline2003}
\\ 
\hline 
Patriot & US-centric, including groups such as ``Birthers'', militia, anti-government, anti-immigration, opposition to financial system.
Some of these themes are not exclusive to ER. & \cite{SPLCIdeology2013}
\\
\hline 
Political Party & Primarily European parties such as the BNP, FP{\"O}, Jobbik, NPD, PVV, Swedish Democrats, UKIP etc. Many of these parties
are also categorized as Populist. & \cite{ZwischenPropaganda2011,doi:10.1080/13691180802158482,GoodwinRadicalRight2012} 
\\ 
\hline 
Populist & Broader category that includes various themes such as anti-EU, anti-establishment, anti-state/government, anti-immigration (as
with Patriot, some of these are not exclusive to ER). Although some disagreement about this category exists \cite{MarlierePopulism2013}, we
have used it as it has proved convenient for categorizing certain groups that span multiple themes.
& \cite{BartlettDigitalPopulism2011,MuddePopulistRadicalRight2007} \\
\hline
Revisionist & References to Holocaust/WWII denial. Closely associated with Neo-Nazi. & \cite{Burris2000,GerstenfeldHateOnline2003,JCC4:JCC410} \\ \hline 
Street Movement & Groups such as the EDL, Autonome Nationalisten, Spreelichter, Anti-Antifa etc. &
\cite{ZwischenPropaganda2011,GoodwinRadicalRight2012}
\\
\hline 
White Nationalist & References to white nationalism and supremacism, also used to characterize political parties such as the BNP or Jobbik.
& \cite{Burris2000,GerstenfeldHateOnline2003,JCC4:JCC410}
\\
\hline
\end{tabular}
\end{center}
\vskip -0.8em
\label{tab:extcategories}
\end{table*}

These channel
documents were transformed to log-based TF-IDF vectors, and subsequently normalized to unit length. However, in an attempt
to reduce the term dimensionality found in the combination of seed and related channel documents,
we generated the TF-IDF vectors in two stages.
The first stage constructed vectors for the seed documents, from which a reduced seed term vocabulary was derived.
In the second stage, vectors for the related documents were then constructed using the seed vocabulary.
In both stages, short documents containing $<10$ terms were excluded. 
We use $m_s$ and $m_r$ to refer to the number of seed and related channel document vectors in a particular data set.

Topic modelling is concerned with the discovery of latent semantic structure or topics within a text corpus, which can be derived from
co-occurrences of words and documents \cite{steyvers2006probabilistic}. 
Popular methods include probabilistic models such as latent Dirichlet allocation (LDA)
\cite{Blei:2003:LDA:944919.944937}, or matrix factorization techniques such as Non-negative matrix factorization (NMF)~\cite{lee99}. 
We initially evaluated both LDA and NMF-based methods with the seed channel document representations described above (the former was
applied to TF vectors). However, as in our previous work, NMF was found to produce the most
readily-interpretable results, which appeared to be due to the tendency of LDA to discover topics that
over-generalized~\cite{ChemuduguntaGeneralSpecificTopic2006,OCallaghanMUSEPostproceedings2013}. We were aware of the presence of smaller groups of channels
associated with multiple languages in both data sets, and opted for specificity rather than generality by applying NMF to the TF-IDF channel vectors. As
we found previously, the IDF component ensured a lower ranking for less discriminating terms, thus leading to the discovery of more specific
topics.

To allow the seed channel vectors to determine the topic basis vectors, we undertook the process in two stages. Firstly, we generate topics
for the seeds:
\begin{enumerate}
    \item Construct an $n \times m_s$ term-document matrix $V_s$, where each column contains a seed channel TF-IDF vector.
    \item NMF is applied to $V_s$ to produce two factors; $W_s$, an $n \times T$ matrix containing topic basis vectors and $H_s$, a $T
    \times m_s$ matrix containing the topic assignments or weights for each seed channel document.
\end{enumerate}
For both data sets, this results in a set of basis vectors consisting of both ER and non-ER topics. The second stage involves producing topic assignments for the related channel vectors:
\begin{enumerate}
    \item Construct an $n \times m_r$ term-document matrix $V_r$ containing the related channel TF-IDF vectors.
    \item Generate a corresponding $T \times m_r$ topic weights matrix $H_r$ by transforming $V_r$ according to the $W_s$ model.
\end{enumerate}
The latter step was achieved with a single outer iteration of the $H$ non-negative least squares sub-problem from the approach by Lin \etal
\cite{Lin:2007:PGM:1288856.1288865}. As the basis vectors matrix $W_s$ was fixed, only one outer iteration was necessary in order to
generate an approximation of a converged $H_r$, thus permitting large related channel matrices to be factorized given the initial NMF operation on a
far smaller seed matrix. 
In both cases, to address the instability introduced by random
initialization in standard NMF, we employed the deterministic NNDSVD initialization method~\cite{Boutsidis:2008:SBI:1324613.1324653}.

\subsection{Topic and Channel Categorization}
\label{categorization}

As discussed earlier in Section \ref{videorecommendation}, some prior work has proposed generic categories for use with YouTube videos and
channels.
However, as these studies have focused on the categorization of mainstream videos, they are not sufficient for the present
analysis where categories specifically associated with the extreme right are required. In Section \ref{ercategorization}, we discuss
prior work that characterized online extreme right activity using a number of proposed categories, but as indicated earlier,
no definitive set of categories is agreed upon in this domain \cite{Blee2010}.
Therefore, we propose a categorization based on various schema found in a selection of academic literature on the extreme right, where
this category selection is particularly suited to the ER videos and channels we have found on YouTube. Some categories are clearly
delineated while others are less distinct, reflecting the complicated ideological make-up and thus fragmented nature of groups and
sub-groups within the extreme right \cite{GerstenfeldHateOnline2003}. In such cases, we have proposed categories that are as specific as possible while also accommodating a number of disparate themes and groups.
Details of the categories employed can be found in Table \ref{tab:extcategories}. 

As both data sets contained various non-ER channels, we also created a corresponding set of non-ER categories consisting of a
selection of the general YouTube categories as of June 2013, in addition to other categories that we deemed appropriate following an inspection of these channels and
associated topics. These non-ER categories were:
\begin{multicols}{2}
\begin{itemize}
	\item Entertainment 
	\item Gaming 
	\item Military 
	\item Music 
	\item News \& Current Affairs 
	\item Politics 
	\item Religion 
	\item Science \& Education 
	\item Sport 
	\item Television 
\end{itemize}
\end{multicols}

Having produced a set of $T$ topics for a data set, we then proceeded to categorize them. For each topic, we manually
inspected the high-ranking topic terms from the corresponding vector in $W_s$, in addition to profiles and uploaded videos for a selection
of seed channels most closely assigned to the topic, according to their weights in $H_s$.
Multiple categories were assigned to topics where necessary, as using a single category per topic would have been too restrictive while
also not reflecting the often multi-faceted nature of most topics that were identified. In many cases, categories for topics were
clearly identifiable, with a separation between ER and non-ER categories. For example, an \textit{English Defence League} (EDL) topic was
categorized as Anti-Islam and Street Movement, while a topic having high-ranking terms such as ``guitar'' and ``band''
was categorized as Music. For certain topics, this separation was more ambiguous, where the channels assigned to a particular topic
consisted of a mixture of both ER and non-ER channels. A combination of both ER and non-ER categories were assigned in such cases.

\begin{figure}[!h]
	\begin{center}
        \begin{subfigure}{0.45\textwidth}
                \centering
                \includegraphics[scale=0.69]{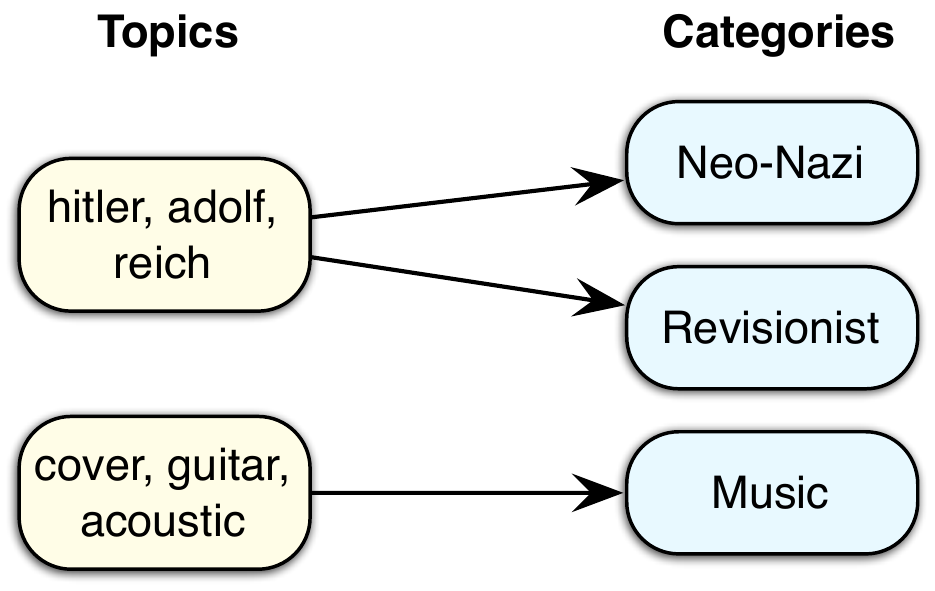}
		      	\vskip 1.0em
                \caption{Categories assigned to topics.}
                \label{fig:categorizetopics}
        \end{subfigure}
      	\vskip 2.0em
		\begin{subfigure}{0.45\textwidth}
                \centering
                \includegraphics[scale=0.69]{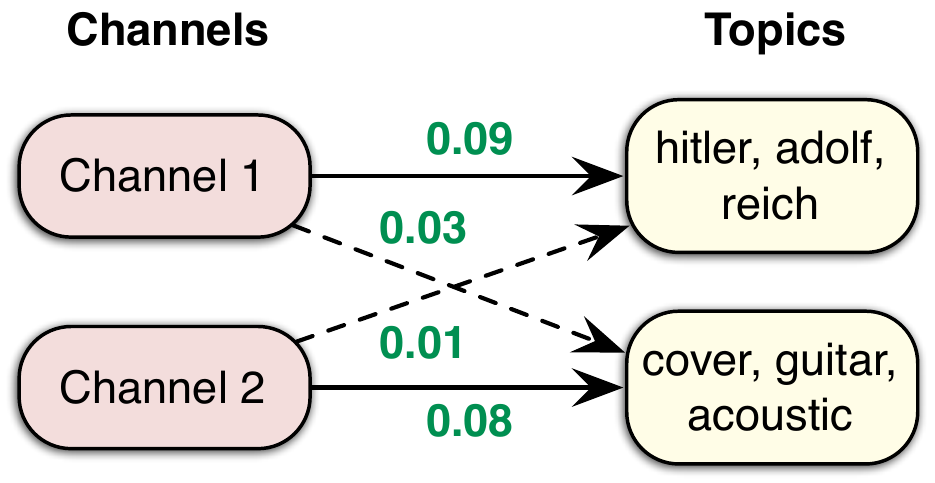}
		      	\vskip 1.0em
                \caption{Channel $H$ weights for topics.}
                \label{fig:userstopics}
        \end{subfigure}    
      	\vskip 2.0em
        \begin{subfigure}{0.45\textwidth}
                \centering
                \includegraphics[scale=0.69]{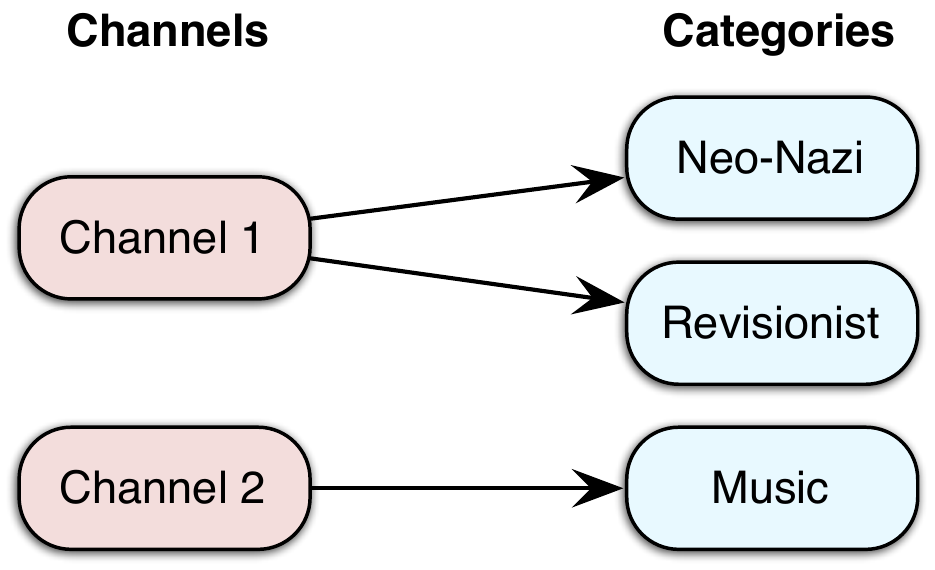}
		      	\vskip 1.0em
                \caption{Channels categorized using topics where $H \ge w$ (in this example, $w=0.05$). }
                \label{fig:categorizeusers}
        \end{subfigure}
	\end{center}
	\vskip -0.8em
	\caption{Topic and Channel categorization process.}
	\label{fig:categorizationprocess}
\end{figure}

The set of categorized topics was then used to label both seed and related channels, using the channel topic assignment weights found
in the $H$ matrices. This supports the potential assignment of multiple categories to a single channel. To achieve this, it was first necessary to determine a
weight threshold $w$ for selecting discriminating topics to be used in the categorization of a particular channel. Using a range of values
for $w$, we calculated $\mu$, the mean number of topics per channel with $H$ weight $\ge w$, and selected the value for $w$ where
$\mu= \sim1$; the minimum number of topics required for categorization. As we performed NMF in two stages to produce $H_s$ and $H_r$
for seed and related channels respectively, two corresponding $w_s$ and $w_r$ thresholds were generated due to the 
dependency of the $H$ weights' upper bound on the input matrices $V$.
Using these thresholds, each channel was categorized as follows:
\begin{enumerate}
    \item Select all topics where the channel's corresponding row weights in $H$ are $\ge$ $w$.
    \item Add each selected topic weight to the totals maintained for the topic's corresponding 
 categories.
    \item Rank these categories in descending order based on their totals across all selected topics, and assign the highest ranked
    category to the channel. Multiple categories are assigned in the event of a tie.
\end{enumerate}
A simple example illustrating the complete process of mapping of channels to categories is shown in Figure \ref{fig:categorizationprocess}.
We found that a certain number of channels, both seed and related, had a flat profile with relatively low weights ($< w$) for all topics;
these could be considered as \textit{grey sheep}.
Further analysis found this to be due to factors such as the original documents being short or containing few unique terms. As we were unable to reliably categorize such
channels, they were excluded from all subsequent analysis. Separately,
although the NMF process often identified topics with high-ranking discriminating terms in languages other than that of the data set, a
small number of topics with less discriminating general language terms were also found. 
As it would have been difficult to distinguish between ER and non-ER channels closely associated with these topics, these were also excluded. 
Further details on the $w$ thresholds and numbers of excluded channels are provided in Section
\ref{analysis}.

\section{The ER Filter Bubble}
\label{analysis}

In this section, we discuss the filter bubble analysis of the English and German language YouTube data sets. 
The experimental methodology is as follows:

\begin{enumerate}
    \item Generate an aggregated ranking of related channels for each seed channel.
    \item Generate TF-IDF channel document vectors, and identify topics using NMF.
    \item Categorize the identified topics according to the set defined in Table \ref{tab:extcategories}.
    \item Categorize the channels based on their topic weights in $H$.
    \item Investigate whether an extreme right filter bubble is present.
\end{enumerate}
For Step 5 above, we define an extreme right filter bubble in terms of the extent to which the related channels for a particular
extreme right seed channel also feature extreme right content. 
It has been shown that the position of a video in a related video list plays a critical role in the click through rate \cite{Zhou:2010:IYR:1879141.1879193}.
Therefore, we investigate the presence of a filter bubble using the top $k$ ranked related channels, with increasing values of $k \in
[1,3,5,7,10]$, as follows.

For each ER seed channel:

\begin{enumerate}
       \item Select the top $k$ ranked related channels, filtering any excluded channels as defined in Section \ref{categorization}. Seed
        channels with no remaining related channels following filtering are not considered for rank $k$.
      \item Calculate the total proportion of each category assigned to the $\le k$ related channels.
\end{enumerate}
Then, for each ER category:
\begin{enumerate}
      \item Select all seed channels to which the category has been assigned.
      \item Calculate the mean proportion of each related category associated with these seed channels.
\end{enumerate} 

We consider a filter bubble to exist for a particular ER category when its highest ranking related categories, in terms of their mean
proportions, are also ER categories.

\subsection{English language categories}
\label{englishanalysis}

From the total number of channels in the English language data set, we generated 24,611 seed and 
1,376,924 related channel documents, using a corresponding seed-based vocabulary of 39,492 terms. The TF-IDF matrices $V_s$ and $V_r$ were
then generated for seed and related documents respectively, and topics were identified by applying NMF to $V_s$. To determine the
number of topics $T$, we experimented with values of $T$ in $[10,100]$ to produce topics that were as specific as
possible, given prior knowledge of the presence of smaller groups of channels associated with multiple languages within both data sets.
This led to the selection of $T = 80$, as larger values resulted in topic splits rather than the emergence of unseen topics. Of these,
27 ER topics (33.75\%), 39 non-ER topics (48.75\%), 8 topics that were a combination of ER and non-ER categories (10\%), and 6
topics based on general terms of a separate language (7.5\%) were found.

\begin{figure}[!h]
    \centering
    \includegraphics[scale=0.49]{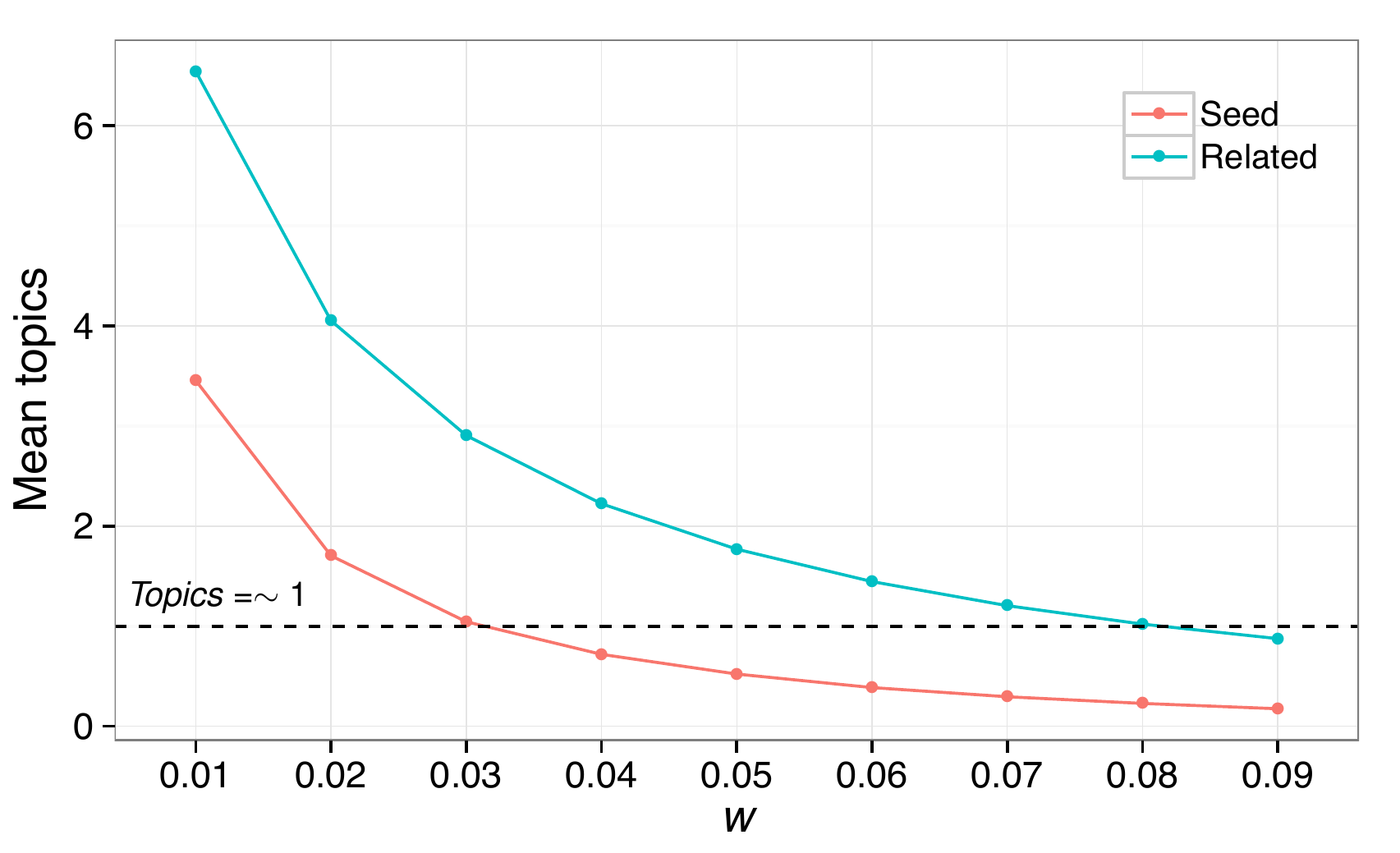}
	\vskip -0.8em
	\caption{English mean topics per channel at $w_s$, $w_r$ in $[0.01,0.09]$. $w_s$ and $w_r$ were set to 0.03 and 0.08 respectively.}
	\label{fig:englishmeantopics}
\end{figure}

These 80 topics were then categorized according to the set defined in Table \ref{tab:extcategories}, which permitted the subsequent
categorization of the seed and related channels using their corresponding $H_s$ and $H_r$ topic weights. As described in Section
\ref{categorization}, we first generated the weights thresholds $w_s$ and $w_r$. Figure \ref{fig:englishmeantopics} contains the mean number
 of topics per channel $\mu$ with $H$ weight $\ge w$, for values of $w$ in $[0.01, 0.09]$. Using $\mu = \sim 1$ as the selection criterion resulted in values of 0.03 and 0.08 for $w_s$ and $w_r$
respectively. Using these, we excluded 8,225 (33.42\%) seed and 482,226 (35.66\%) related grey sheep channels that could not
be categorized. 
Related channels ranked at $k > 10$ were also excluded, and all non-ER seed channels were removed from the candidate seed set. 
The remaining 6,573 ER seed and their 22,980 related channels were used to calculate the mean related category proportions for each ER
category.

\begin{figure}[!t]
	\begin{center}
        \begin{subfigure}{0.49\textwidth}
                \includegraphics[scale=0.49]{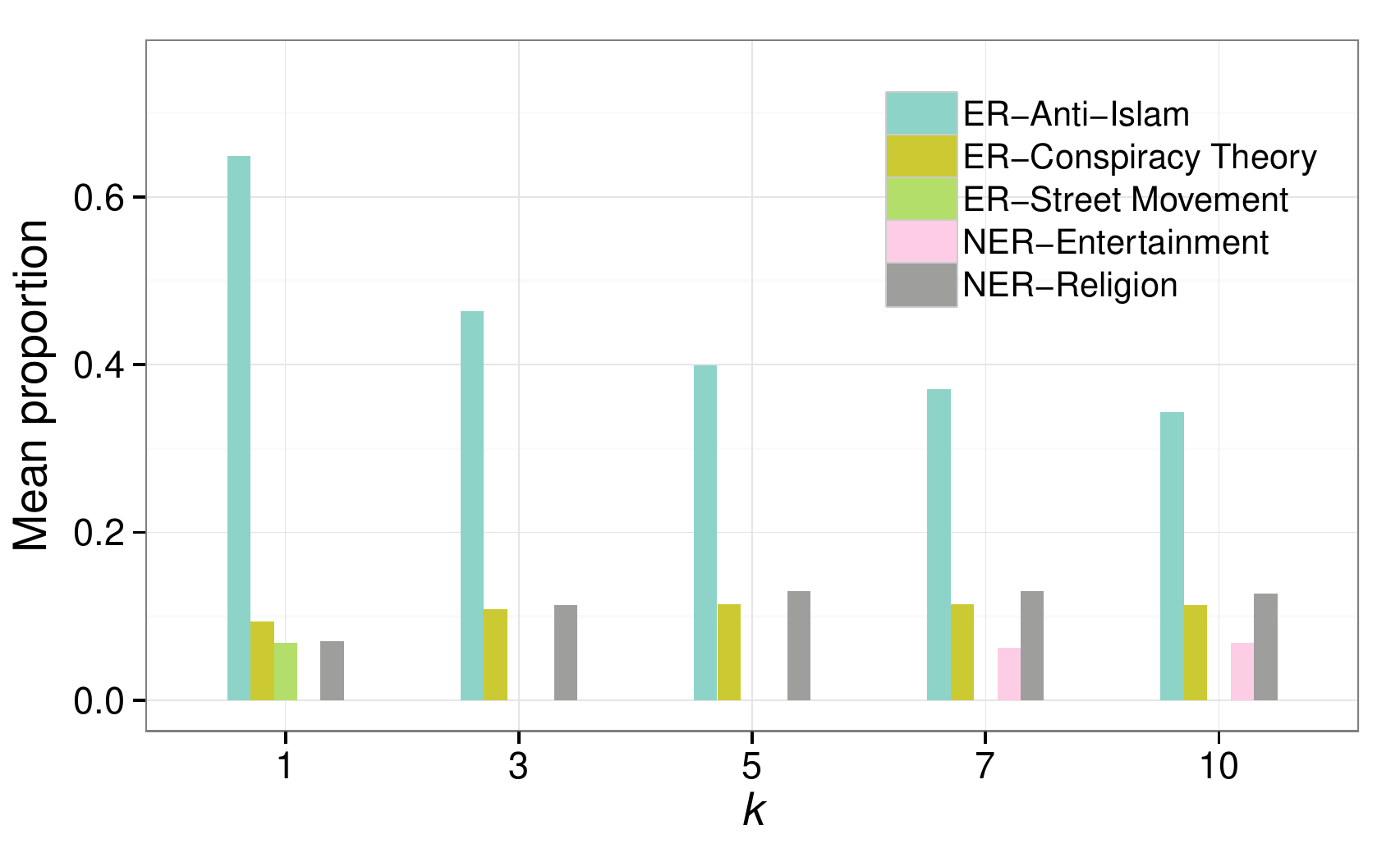}
                \caption{Anti-Islam}
                \label{fig:englishantiislam}
        \end{subfigure}
       	\qquad
		\begin{subfigure}{0.49\textwidth}
                \includegraphics[scale=0.49]{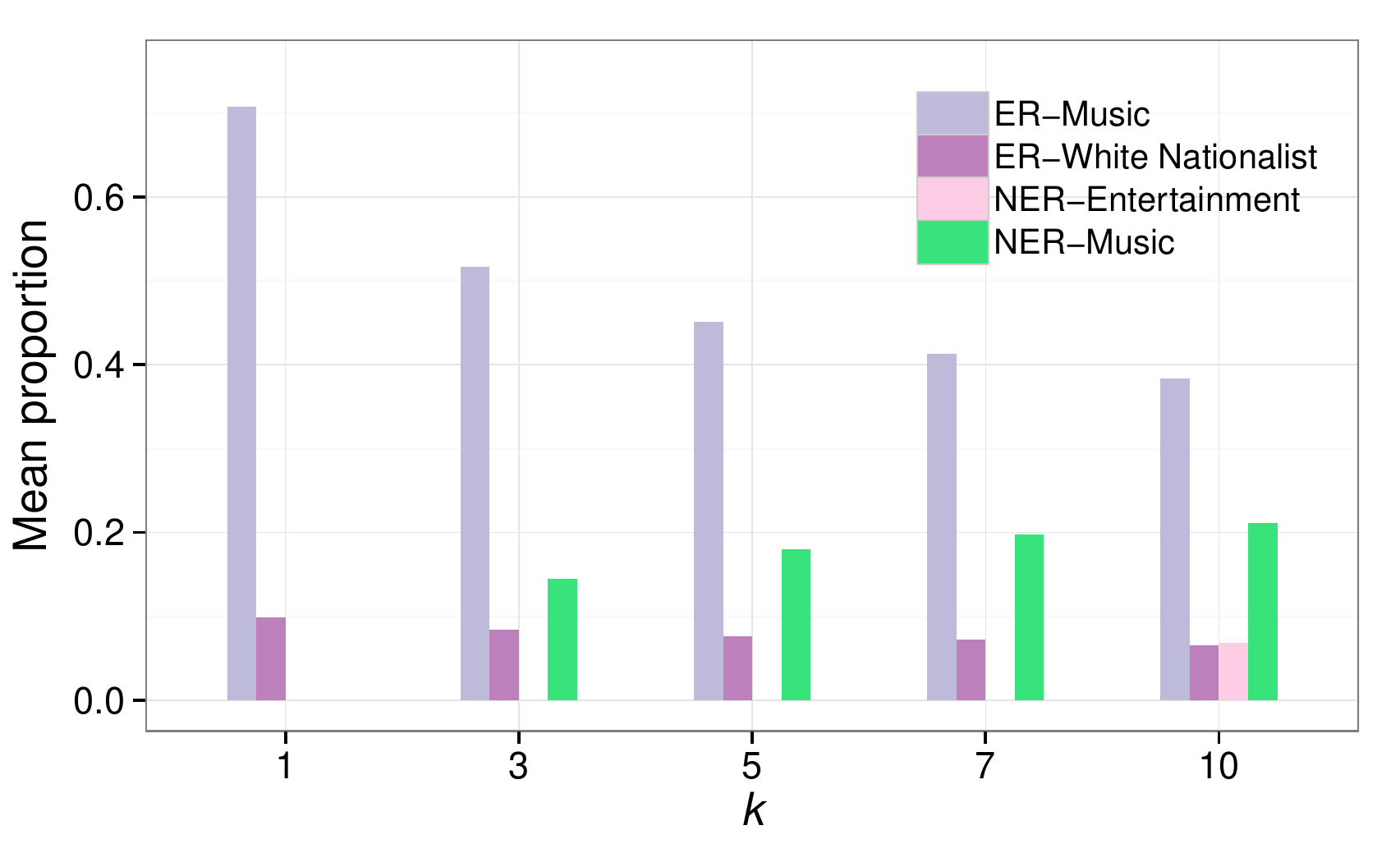}
                \caption{Music}
                \label{fig:englishmusic}
        \end{subfigure}    
       	\qquad
		\begin{subfigure}{0.49\textwidth}
                \includegraphics[scale=0.49]{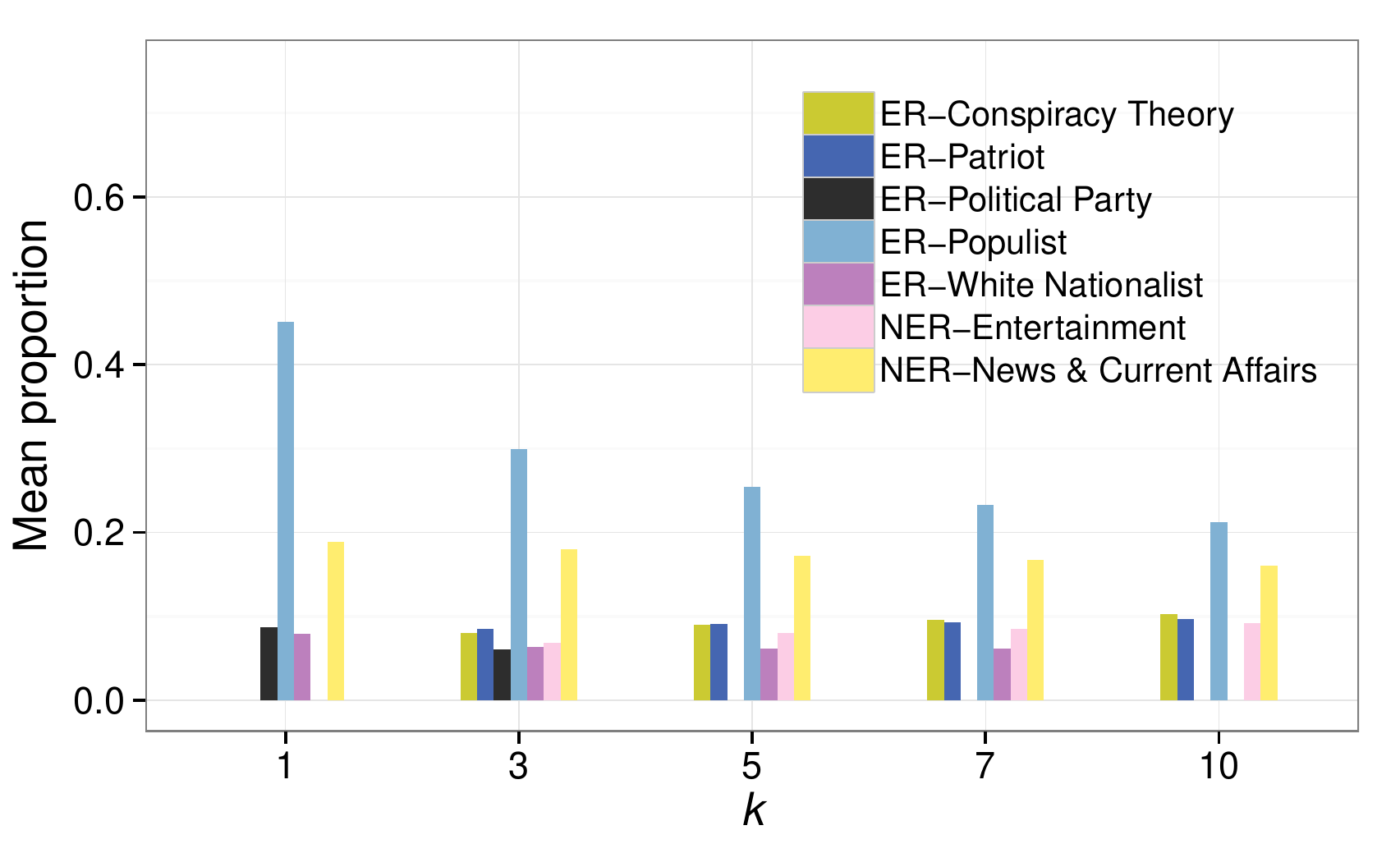}
                \caption{Populist}
                \label{fig:englishpopulist}
        \end{subfigure}    
	\end{center}
	\vskip -1.2em
 	\caption{English mean category proportions of top $k$ ranked related channels ($k \in [1,3,5,7,10]$), for selected seed ER
	categories.}
	\vskip -0.8em
	\label{fig:englishrelatedproportions}
\end{figure}

Figure \ref{fig:englishrelatedproportions} contains plots for three ER categories that we have selected for more detailed analysis,
where ER and non-ER categories have been prefixed with \textit{ER-} and \textit{NER-} respectively. To assist visual interpretation, we
have omitted any weakly related categories whose mean proportion $< 0.06$ for a particular $k$ ranking. From inspecting these plots, two
initial observations can be made; (1) the seed category is the dominant related category for all values of $k$, and (2) although related category diversity
increases at lower $k$ rankings with the introduction of certain non-ER categories, ER categories consistently have the strongest presence.
In the case of Anti-Islam seed channels, it appears that the top ranked related channels ($k=1$) are mostly affiliated with various groups
from the UK
and USA. Street Movement related channels at this rank are associated with the \textit{English Defence League} (EDL), a movement opposed to
the alleged spread of radical Islamism within the UK \cite{GoodwinRadicalRight2012}. Channels from various international individuals and
groups that often describe themselves as ``counter-Jihad'' can also be observed \cite{GoodwinEDLCounterJihad2013}.
The Conspiracy Theory and non-ER Religion categories appear to be associated with channels based in the USA, where the dominance of these
categories at lower rankings (excluding the seed category) suggests that the channels become progressively more US-centric. 
 
The ER Music seed channels usually upload video and audio recordings of high-profile groups associated with the extreme
right. For example, content from bands such as Skrewdriver (UK) or Landser (Germany) can be found, along with other groups from genres such
as Oi!, RAC (Rock Against Communism) and NSBM (National Socialist Black Metal) \cite{ZwischenPropaganda2011,BrownNaziRock2004}. Given this, the
consistent presence of the White Nationalist related category would appear logical. At the same time, we also observe that non-ER Music
becomes more evident as $k$ increases, perhaps reflecting the overlap between music genres. For example, someone who is a
fan of NSBM is often a fan of other metal music that would not be categorized as ER. For Populist seed categories, the related categories
generally appear more diverse. Channels affiliated with political parties can be observed, including the Eurosceptic
\textit{United Kingdom Independence Party} (UKIP) or the \textit{British National Party} (BNP), where the latter is also
considered as White Nationalist \cite{BartlettDigitalPopulism2011,MuddePopulistRadicalRight2007}. Opposition to establishment
organizations such as the EU may be a link to similar opposition within the Patriot and Conspiracy Theory related categories that are also
present \cite{SPLCIdeology2013}, while also explaining the presence of the non-ER News \& Current Affairs category. It
should also be mentioned that our definition of Populist is broad and spans multiple themes (Table \ref{tab:extcategories}), where a certain
amount of disagreement about this category exists \cite{MarlierePopulism2013}.

\subsection{German language categories}
\label{germananalysis}

\begin{figure}[!h]
    \centering
    \includegraphics[scale=0.49]{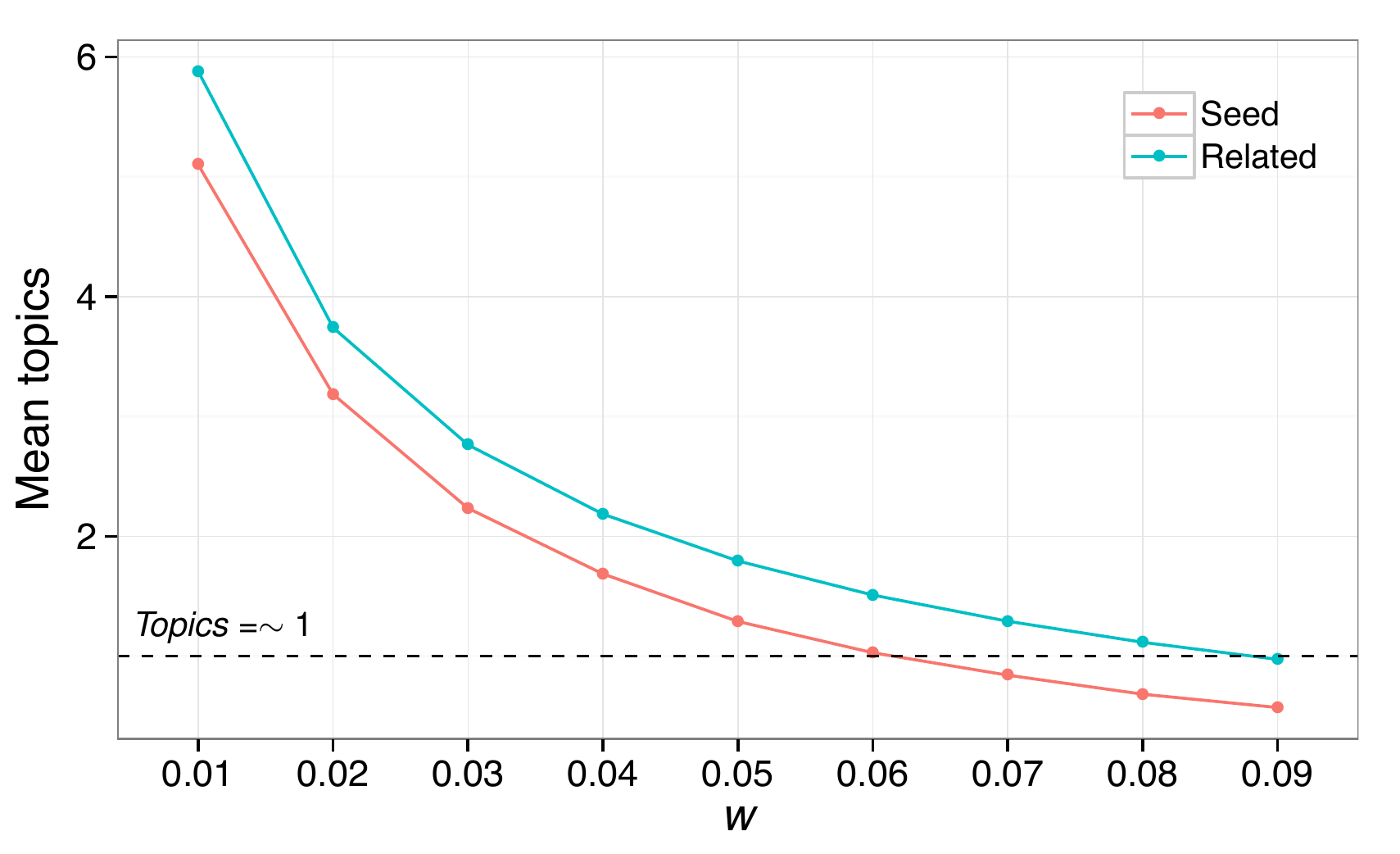}
	\vskip -0.8em
	\caption{German mean topics per channel at $w_s$, $w_r$ in $[0.01,0.09]$. $w_s$ and $w_r$ were set to 0.06 and 0.087 respectively.}
	\label{fig:germanmeantopics}
\end{figure}

A total of 2,766 seed and 177,868 related channel documents were generated from the German language data set.
Topics were then identified by applying NMF to the corresponding TF-IDF matrices. As before, we experimented with values of $T$ in
$[10,100]$, and selected $T = 60$ given a similar observation of redundant topic splits for larger values of $T$. Of these, 33 ER topics
(55\%), 20 non-ER topics (33.33\%), 2 topics that were a combination of ER and non-ER categories (3.33\%), and 5 topics based on general
terms of a separate language (8.33\%) were found.
Having categorized these topics, we calculated the mean number of topics per channel $\mu$ with $H$ weight $\ge w$, for values of $w$ in $[0.01, 0.09]$, as seen in Figure \ref{fig:germanmeantopics}.
Values of 0.06 and 0.087 were found for $w_s$ and $w_r$
respectively at $\mu = \sim 1$. We excluded 785 (28.38\%) seed and 56,565 (31.8\%) related grey sheep channels
that could not be categorized, in addition to  related channels ranked at $k > 10$ and non-ER seed channels.
The remaining 1,123 ER seed and 4,973 related (ER and non-ER, $k \le 10$) channels were used to calculate the mean related category
proportions for each ER category.

\begin{figure}[!t]
	\begin{center}
        \begin{subfigure}{0.49\textwidth}
                \includegraphics[scale=0.49]{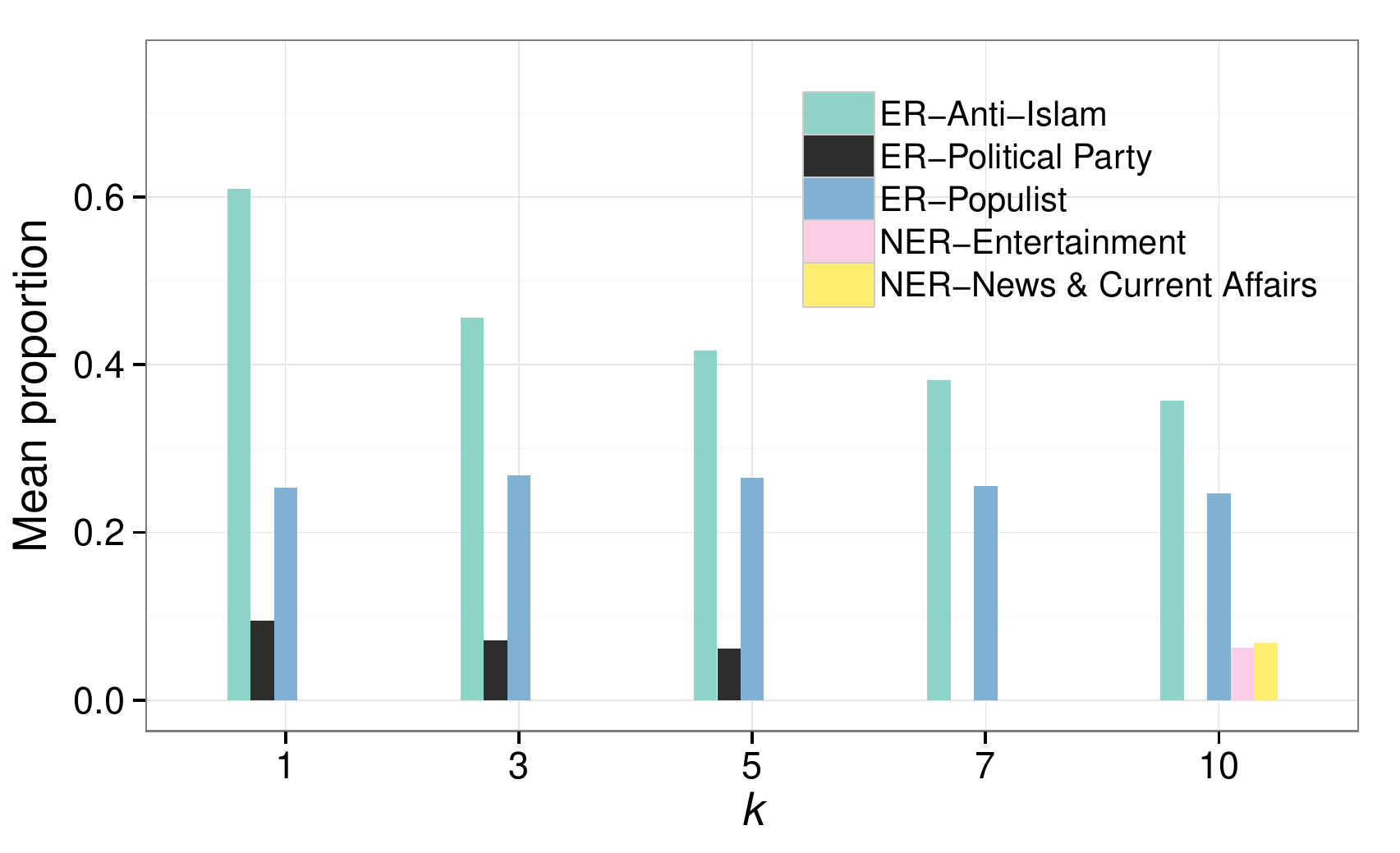}
                \caption{Anti-Islam}
                \label{fig:germanantiislam}
        \end{subfigure}
		\begin{subfigure}{0.49\textwidth}
                \includegraphics[scale=0.49]{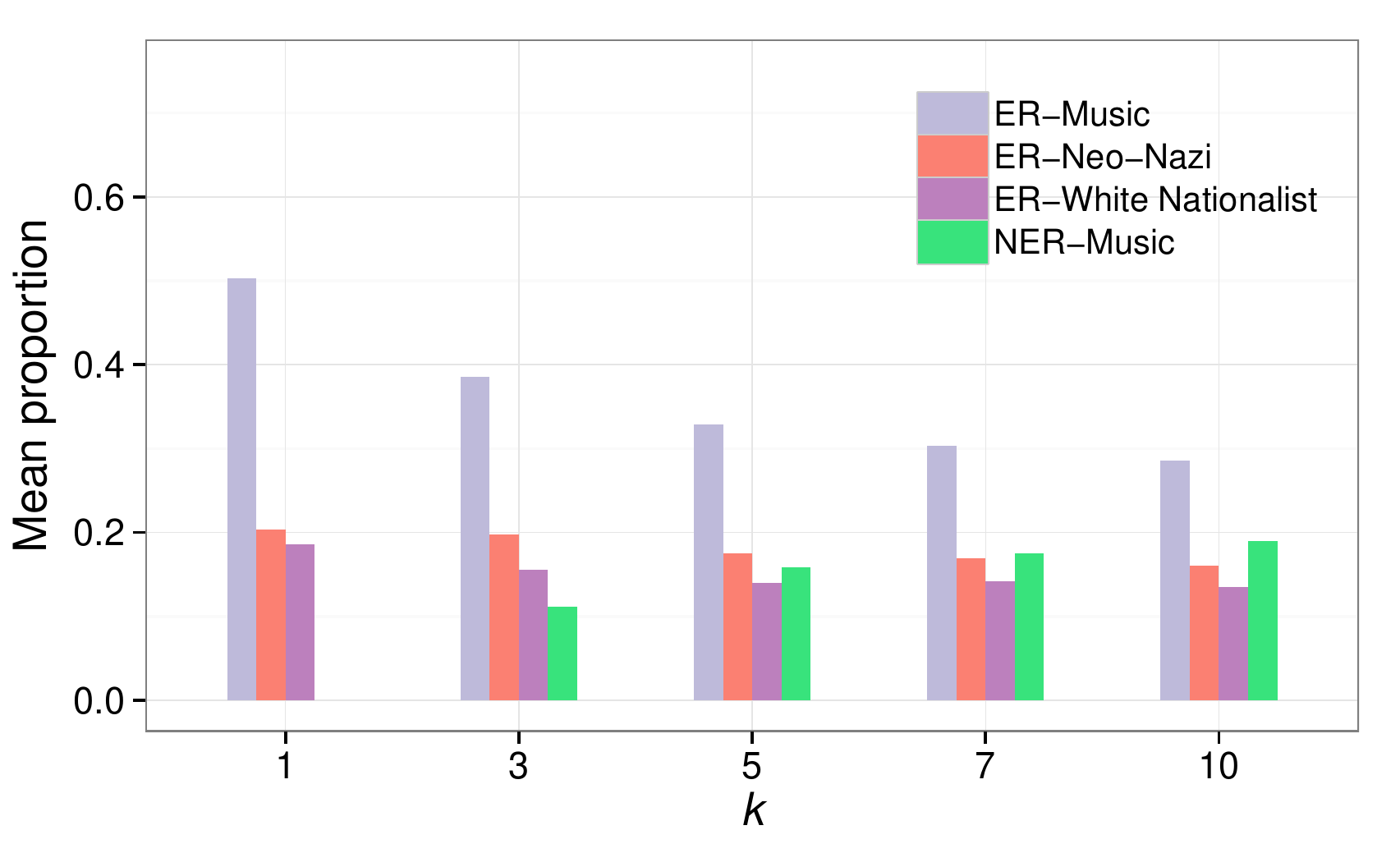}
                \caption{Music}
                \label{fig:germanmusic}
        \end{subfigure}    
        \begin{subfigure}{0.49\textwidth}
                \includegraphics[scale=0.49]{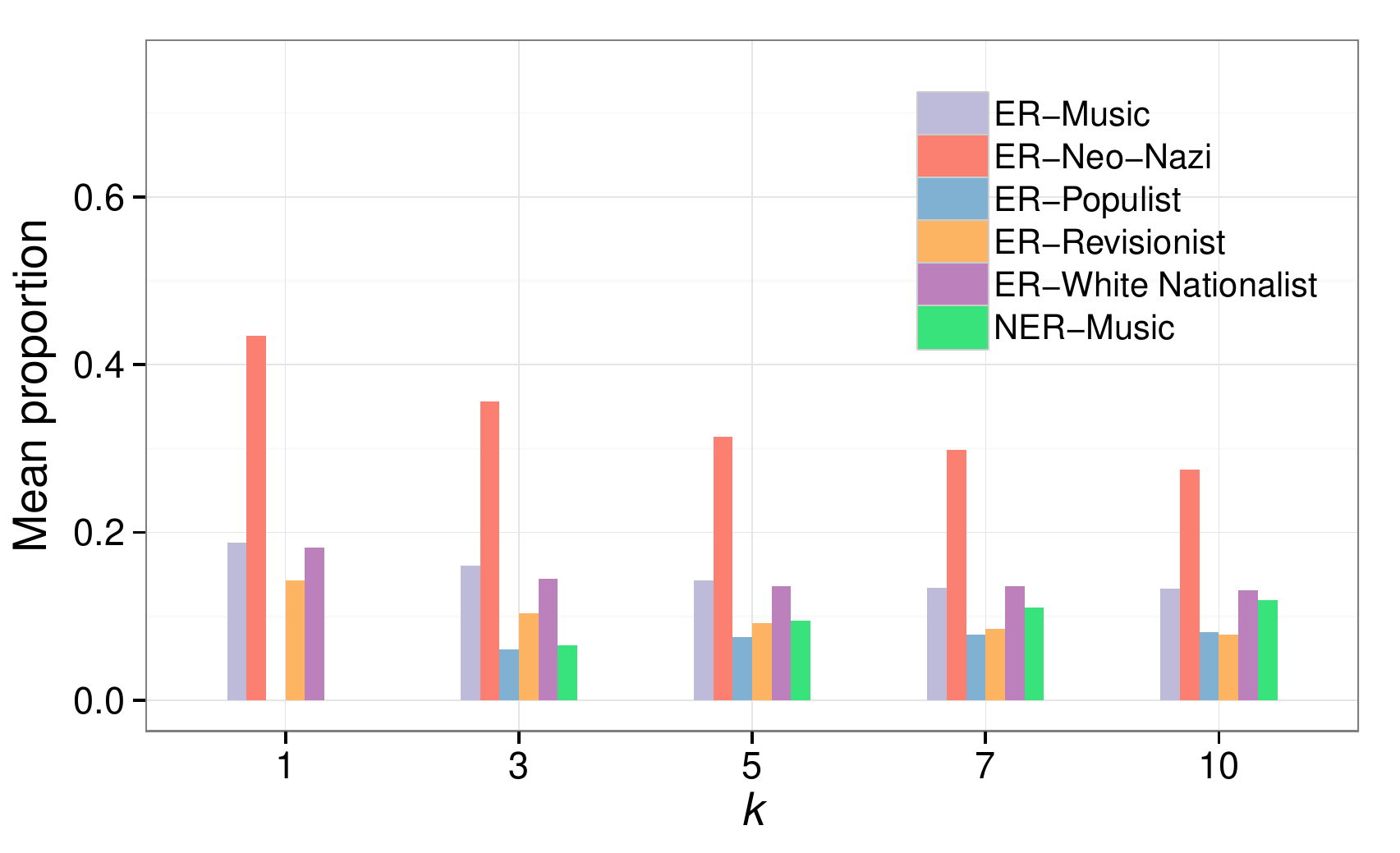}
                \caption{Neo-Nazi}
                \label{fig:germanneonazi}
        \end{subfigure}
	\end{center}
	\caption{German mean category proportions of top $k$ ranked related channels ($k \in [1,3,5,7,10]$), for selected seed ER
	categories.}
	\label{fig:germanrelatedproportions}
\end{figure}

Figure \ref{fig:germanrelatedproportions} contains plots for three ER categories that we have selected for in-depth analysis.
As seen in Figure \ref{fig:englishrelatedproportions}, the seed category is the dominant related category for all values
of $k$, and the ER related category presence is consistently stronger than that of the non-ER categories, notwithstanding the increase in
diversity.
The Populist and Political Party related categories are prominent for Anti-Islam, given the inclusion of channels affiliated with parties
such as the National Democratic Party of Germany (NPD), the Pro-Bewegung collective, and the Freedom Party of Austria (FP{\"O}); all strong
opponents of immigration, particularly by Muslims \cite{ZwischenPropaganda2011,BartlettDigitalPopulism2011, GoodwinEDLCounterJihad2013}. 

This data set also features many ER Music seed
channels that upload recordings of high-profile groups, with the main difference being the prominence of the Neo-Nazi related category for
all rankings. These recordings and videos, along with other non-music videos uploaded by these channels, often feature recognizable
Nazi imagery. Separately, channels that upload videos associated with groups that have alleged ER ties, for example, B{\"o}hse Onkelz or
Frei.Wild \cite{ZwischenPropaganda2011}, may explain in part the presence of the non-ER Music category, given the mainstream success of
these groups. This may also be explained by material associated with hip-hop groups such as ``n'Socialist Soundsystem'', which provide an
alternative to traditional ER music based on rock and folk.
The close relationship with Music is also present for the Neo-Nazi seed category, although further related diversity can be observed.
Seed channels featuring footage of German participation in WWII, including speeches by high-ranking members of the Nazi party, are likely to
be the source of the White Nationalist and Revisionist related categories. We can safely assume that the prominence of Music is
responsible for the appearance of its non-ER counterpart here.

\begin{figure*}
	\begin{center}
		\hskip -1.2em
        \begin{subfigure}[b]{0.47\textwidth}
                \centering
                \includegraphics[scale=0.49]{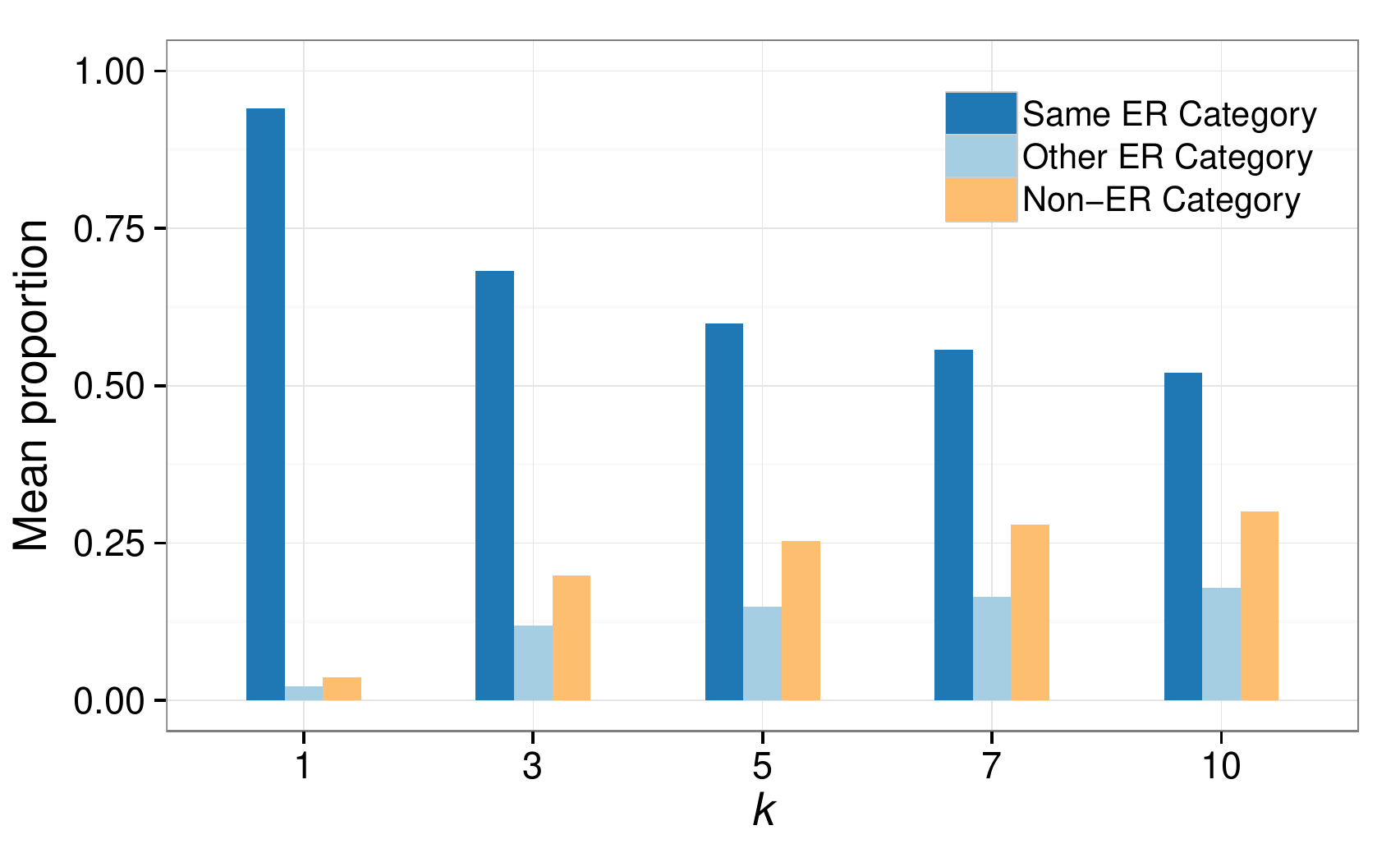}
                \caption{English}
                \label{fig:englishsummary}
        \end{subfigure}
       	\qquad
       	\hskip 1.5em
		\begin{subfigure}[b]{0.47\textwidth}
                \includegraphics[scale=0.49]{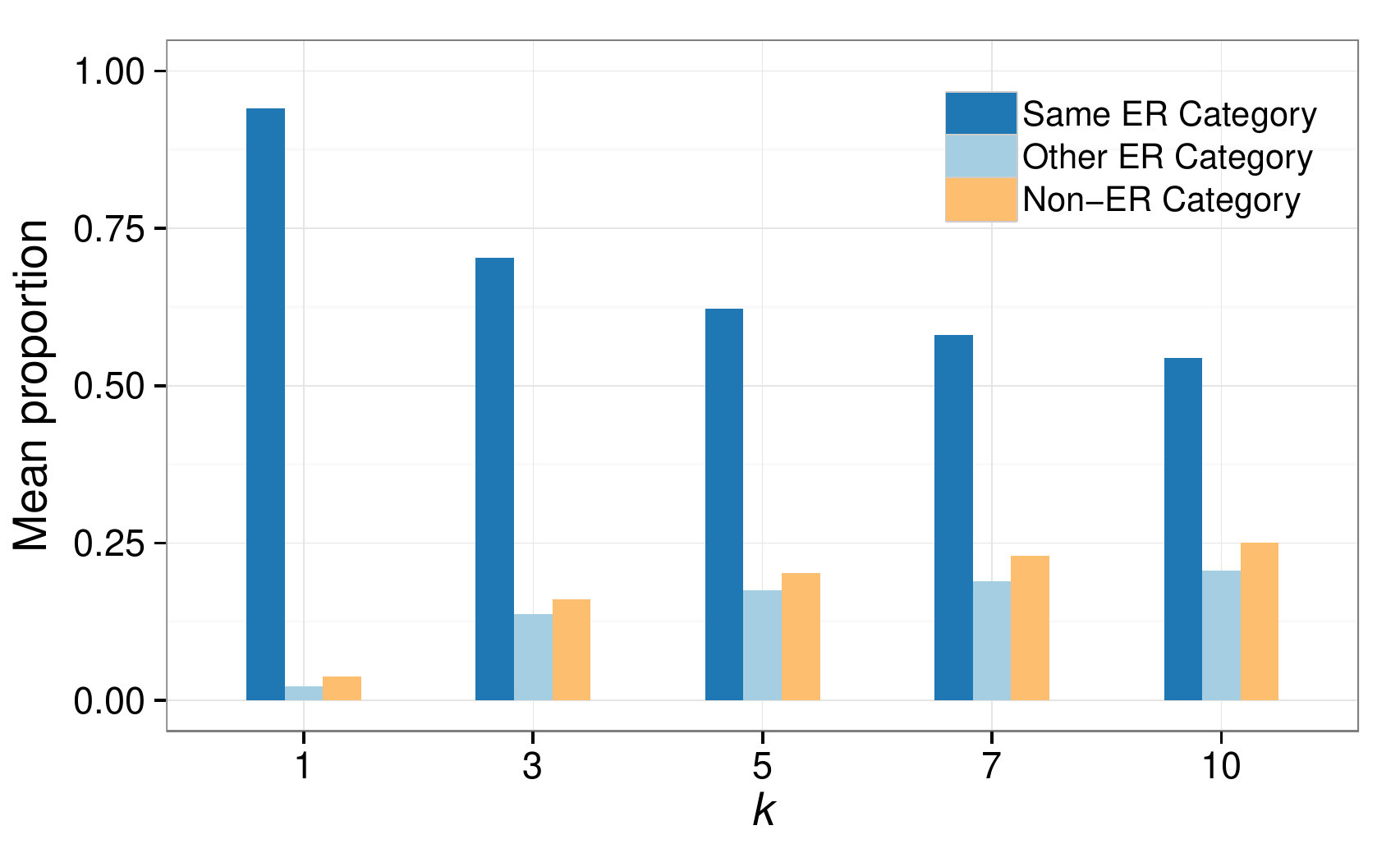}
                \caption{German}
                \label{fig:germansummary}
        \end{subfigure}    
	\end{center}
	\caption{Aggregated mean category proportions of top $k$ ranked related channels ($k \in [1,3,5,7,10]$), for seed ER
	categories.}
	\label{fig:summary}
\end{figure*}

\subsection{Discussion}

We conclude our filter bubble analysis by measuring the mean proportions for the seed ER categories as a whole, where the possible
aggregated related categories were (1) the same ER category as that of the seed, (2) a different ER category, or (3) a non-ER category. The
results for both data sets can be found in Figure \ref{fig:summary}. 
As with the individual seed categories, an ER filter bubble is also clearly identifiable at the aggregate level.
Although the increase in diversity for lower $k$ rankings introduces a certain proportion of non-ER categories, this is always outweighed
by ER categories, where the seed ER category remains dominant for all values of $k$. These findings would appear to contrast those of
certain prior work where greater related video diversity was observed 
\cite{Roy:2012:SCT:2393347.2393437,Zhou:2010:IYR:1879141.1879193}. Although we have analyzed related channels
rather than individual videos, we might have expected to also find this behaviour at both levels. However, it would appear
this is not always true, at the very least in the case of ER channels.

The data retrieval process described in Section \ref{data} involved following related video links for only one step removed from the
corresponding seed video; no further related retrieval was performed for related videos themselves. Of the ER seed
channels used in the filter bubble analysis in Sections \ref{englishanalysis} and \ref{germananalysis}, we
identified those that appeared in top $k$ related rankings of other ER seed channels, for $k \le 10$; 6,186 (94\%) and
1,056 (94\%)
such channels respectively for the English and German language data sets.
These high percentages
allude to the presence of cycles within the related channel graph, where retrieving additional data by following related videos for multiple steps may have been somewhat redundant.
They also further emphasize the existence of the filter bubble. 

Separately, it might be argued that our findings merely confirm that YouTube's related
video recommendation process is working correctly. However, this is precisely what we were trying to quantify, in terms of the categories
of extreme right content to which a user may be exposed following a short series of clicks. As a consequence, users viewing this content may rarely be
presented with an alternative perspective.

\section{Conclusions and Future Work}
\label{conclusions}

YouTube's position as the most popular video sharing platform has resulted in it playing an important role in the online strategy
of the extreme right, where it is used to host associated content such as music and other propaganda. We have proposed a
set of categories that may be applied to this YouTube content, based on a review of those found in existing academic studies of the extreme
right's ideological make-up.
Using an NMF-based topic modelling approach, 
we have categorized channels according to this proposed set, permitting the assignment of multiple categories per channel where necessary.
This categorization has helped us to identify the existence of an extreme right filter bubble, in terms of the extent to which related
channels, determined by the videos recommended by YouTube, also belong to extreme right categories. Despite the increased diversity
observed for lower related rankings, this filter bubble maintains a constant presence. The influence of related rankings on click through
rate \cite{Zhou:2010:IYR:1879141.1879193}, coupled with the fact that the YouTube channels in this analysis originated from links posted by
extreme right Twitter accounts, would suggest that it is possible for a user to be immersed in this content following a short series of
clicks.

In future work, we would like to experiment with alternative channel representations that address the existence of noise found with
certain channels and associated topics at lower related rankings. We would also like to perform a longer temporal analysis in an attempt to
investigate any correlation between changes in this filter bubble and extreme right activity across multiple social media platforms. Separately, although we have used the methodology presented
in this paper to study extreme right channels, it may be interesting to investigate its application to other types of extremist political
content found on YouTube.

\section{Acknowledgments}
%\emph{Anonymized}
This research was supported by 2CENTRE, the EU funded Cybercrime Centres of Excellence Network and Science Foundation Ireland (SFI) under
Grant Number SFI/12/RC/2289.
% Balancing columns in a ref list is a bit of a pain because you
% either use a hack like flushend or balance, or manually insert
% a column break.  http://www.tex.ac.uk/cgi-bin/texfaq2html?label=balance
% multicols doesn't work because we're already in two-column mode,
% and flushend isn't awesome, so I choose balance.  See this
% for more info: http://cs.brown.edu/system/software/latex/doc/balance.pdf
%
% Note that in a perfect world balance wants to be in the first
% column of the last page.
%
% If balance doesn't work for you, you can remove that and
% hard-code a column break into the bbl file right before you
% submit:
%
% http://stackoverflow.com/questions/2149854/how-to-manually-equalize-columns-
% in-an-ieee-paper-if-using-bibtex
%
% Or, just remove \balance and give up on balancing the last page.
%
\balance
 
% If you want to use smaller typesetting for the reference list,
% uncomment the following line:
% \small
\bibliographystyle{acm-sigchi}
\bibliography{main}
\end{document}